# Weak Antilocalization and Linear Magnetoresistance in The Surface State of SmB$_6$

S. Thomas[1]†, D.J. Kim[1]†, S. B. Chung[2], T. Grant[1], Z. Fisk[1] and Jing Xia[1]*

[1] Dept. of Physics and Astronomy, University of California, Irvine, California 92697, USA

[2] Dept. of Physics and Astronomy, University of California, Los Angeles, California 90095, USA

* Correspondence to: xia.jing@uci.edu

† These authors contributed equally to this work.

Strongly correlated Kondo insulator SmB$_6$ is known for its peculiar low temperature residual conduction, which has recently been demonstrated to arise from a robust metallic surface state, as predicted by the theory of topological Kondo insulator (TKI). Photoemission, quantum oscillation and magnetic doping experiments have provided evidence for the Dirac-like dispersion and topological protection. Questions arise as whether signatures of spin-momentum locking and electron interaction could be resolved in transport measurements. Here we report metallic conduction of surface state down to mK temperatures with saturation behaviors suggestive of Kondo effect. We observe in the surface state the weak-antilocalization (WAL) effect that is in agreement with a spin-momentum locked metallic surface. At larger perpendicular magnetic fields, the surface state exhibits an unusual linear magnetoresistance similar to those found in Bi-based topological insulators and in graphene.

With its heavy f-electron degree of freedom, Kondo insulator (*1*) SmB$_6$ (*2*) behaves as a correlated metal at high temperatures. Below 40 K the bulk of SmB$_6$ becomes insulating with the opening of Kondo energy gap due to the hybridization between conduction electrons and the highly renormalized f-electrons. The theory (*3*) of topological Kondo insulator predicted the existence of a topologically protected surface state (TSS) within this Kondo gap, naturally explaining the mysterious resistance saturation below 4 K (*1*). Recent transport measurements (*4-6*) have confirmed the low temperature surface conduction and the robustness of the surface state (SS). This SS has been demonstrated (*7*) to vanish with a small amount of magnetic impurity but survives larger amount of non-magnetic doping, which is consistent with a TSS protected by time-reversal symmetry. Recent high resolution ARPES (*8-10*) and quantum oscillation (*11*) experiments have provided tentative evidence for the Dirac dispersion of the surface carriers, as expected in a TSS. Furthermore, unlike usual topological SS, first principle calculations (*12, 13*) have predicted three surface Dirac bands residing at Γ and X/Y points, which agrees with ARPES-measured surface electronic structure (*8-10*), although the anticipated spin-momentum locking (*14, 15*) awaits spin-resolved measurements. In this paper we perform transport studies of the SmB$_6$ SS down to 20 mK, searching for transport signatures of spin-momentum locking and electron interaction effects.

Extending our previous work (*6*), we have verified the existence of SS in SmB$_6$ samples down to 20mK with non-local transport and thickness dependent Hall effect measurements. In particular, Fig. 1A demonstrates the non-local measurement from sample S11 with both (100) and (101) surfaces. The local voltage $V_{Local}$ is measured at voltage leads close to the current lead on the (101) surface. And the non-local voltage $V_{NL}$ is measured at voltage leads placed on the (100) side surface at the corner of the sample. The crossover from high temperature bulk conduction to low temperature surface dominated conduction is seen clearly from the temperature dependence of voltage ratio $V_{NL}/V_{Local}$, which increase 4 orders of magnitude as temperature is lowered, and saturates below 4 K down to 20 mK (Fig. 1B). We note that $V_{NL}/V_{Local}$ approaches a large value of 30, which indicates better conduction on the (101) surface than (100) surface. We found both longitudinal resistance $R_{xx}$ and hall slope $R_{xy}/B$ continue their saturation to 20 mK, indicating the metallic surface doesn't localize even at our lowest temperature. A typical example is shown in Fig. 1C, D for a bar shaped sample with exposed (100) surfaces. The low temperature saturation behavior of $R_{xx}$ is however unlike that of a simple metal, but displays logarithmic temperature dependence (Fig. 1D) that we will discuss later in this paper.

Weak antilocalization (WAL) (*16*) is expected in a TSS due to spin-momentum locking (*14, 15*), which causes destructive interference between time-reversed electron paths and lowers the sample resistance. This effect will be destroyed by a time-reversal-symmetry-breaking magnetic field, giving rise to magneto-resistance dip around zero field. Indeed we observe clear WAL effect in $SmB_6$ Samples at low temperatures. Fig. 2A shows the WAL effect at 20 mK of a thin plate $SmB_6$ sample S12B with dimensions $t = 120\ \mu m, w = 1100\ \mu m, and\ L = 3300\ \mu m$. With magnetic field perpendicular to the plate surface ($\theta = 90°$), a sharp resistivity dip occurs around zero field. The effect becomes almost invisible when the field is rotated to be perpendicular to the sidewall ($\theta = 0°$). In the limit of long inelastic scattering time, WAL gives a negative quantum correction to the conductivity described by Hikami-Larkin-Nagaoka (HLN) equation (*17*): $\Delta G = -\alpha\ [e^2/2\ \pi^2 \hbar][\ln(B_0/B) - \Psi(1/2 + B_0/B)]$, where $\Psi(x)$ is digamma function, $B$ is perpendicular magnetic field component and $B_0 = \hbar/(4eL_\phi^2)$ with $L_\phi$ as dephasing length. Each independent conduction channel will contribute $\alpha = 1/2$. In Fig. 2B we fit the $\theta = 90°$ data to HLN equation and found $L_\phi = 1800\ nm$ and $\alpha = 0.92$. At $\theta = 0°$ the smaller side surfaces would contribute to WAL and should result in WAL of a different size: $\alpha' = \alpha \cdot t/w = 0.1$, which agrees reasonably with $\alpha' = 0.15$ that is fitted by fixing $L_\phi = 1800\ nm$. In other samples (Fig. S1) we have found $\alpha$ to be 0.8, 0.95 and 0.97, suggesting a universal value of 1. Taking into account both top and bottom surfaces, one might expect $\alpha = 3$ for a TI with three Dirac bands that have exactly the same $L_\phi$. However, if one of the Dirac bands (e.g. Γ pocket) has a much larger $L_\phi$ than other bands, it will dominate WAL at small fields and make $\alpha$ appear to be 1. Alternatively, if inter-Dirac-band scattering become important compared to scatterings within an individual Dirac band, the three bands would behave effectively as a single channel as far as transport is concerned and will yield $\alpha = 1$. If indeed $SmB_6$ is a TKI with three Dirac bands (*3, 12, 13, 18*), the $\alpha \approx 1$ value indicates either different dephasing lengths among surface bands or the importance of inter-band scattering. Fixing $\alpha = 0.92$ we could fit WAL data at different temperatures (Fig. 3C) and extract the temperature dependence of $L_\phi$ (Fig. 3D), which would be suggestive of dephasing mechanism. Below T = 100 mK, $L_\phi$ scales as $T^{-0.55}$, which agrees with 2D electron-electron scattering. Above 100 mK, a deviation from such a power law suggests the onset of other scattering mechanisms, e.g. from phonons.

The WAL effect would also result in a logarithmic decrease of resistance when temperature is lowered (*16*), a behavior that, like the WAL itself, would be quenched by a magnetic field. This however, is masked by the logarithmic resistance increase we observe from 1K to 100 mK (Fig. 1D) with the rate $dR_{xx}/d(\ln(T))$ almost independent of applied magnetic field (Fig. 3A). For two dimensional metals both interaction-induced Altshuler–Aronov (AA) effect (*16*) and the Kondo effect (*19*) could give rise to logarithmic resistance that survives large magnetic field. The AA effect gives a quantum correction to resistivity (*16*) $\Delta\rho = \rho^2 A \frac{e^2}{\pi h}\ln(T_1/T_2)$ when the temperature changes from $T_1$ to $T_2$, where $\rho$ is resistivity, $A \leq 1$ is a constant, $e$ is electron charge and $h$ is Plank's constant. From 1K to 100 mK, AA effect thus will increase the resistance $R_{xx}$ in sample S12B by at most 0.3 Ω, which won't fully account for the observed 10 Ω increase (Fig. 1D). Furthermore, we found the relative changes of resistivity $\Delta\rho/\rho$ differ among samples by an order of magnitude, yet obey a very similar temperature dependence (Fig. 3B). This is hardly explained by AA correction alone, but is well described phenomenologically by Kondo effect. For spin 1/2 Kondo effect (*20*), the low temperature saturation resistivity is given by $\rho(T=0) = \rho_c + \rho_K$ and the resistivity in the logarithmic regime is given by $\rho(T) = \rho_c + \frac{\rho_K}{2}[1 - 0.47\ln(1.2T/T_K)]$, where $\rho_c$ and $\rho_K$ represent the sizes of background and Kondo resistivities, $T_K$ is the characteristic Kondo temperature. Using sample S12B (Fig. 1D), we could extract these three parameters by fitting in both the saturation and logarithmic regimes, yielding $T_K = 0.6\ K$. In Fig. 3B, we present the normalized resistance of 4 samples from different growth batches. Due to the similar temperature dependence, a universal $T_K = 0.6\ K$ seems to fit for all the samples. We could compare the normalized Kondo resistance $(\rho_{xx} - \rho_c)/\rho_K$ to the universal Kondo behavior calculated from numerical renormalization group calculations (NRG) (*20*). As shown in Fig. 3C, from 20 mK to 1 K, the normalized Kondo resistance from different samples collapse to a single curve, which agrees well with NRG universal curve. The only exception is below 100 mK with sample S12D where both (100) and (101) surfaces exist on the measurement current path. In a perfect $SmB_6$ crystal, the Kondo effect from localized f-electrons and conduction electrons ceases to exist at low enough temperatures with the opening of the hybridization gap and due to coherent Kondo resonance (*1*). However, when unavoidable trace amount of non-magnetic impurity substitutes Sm atoms near the surface, the resulting "Kondo holes" will give rise to the Kondo effect in the ungapped metallic surface state. Since $T_K$ is determined by the strength of surface Kondo screening, a "universal" $T_K$ is indeed expected, as we found experimentally. The relative size of resistivity change $\Delta\rho/\rho$, however, depends on several sample-specific factors including surface mobility and impurity concentration.

Metals with a closed Fermi surface and a principal charge carrier usually have positive magnetoresistance (MR) that is quadratic with magnetic field. In $SmB_6$ at relative high temperatures when both bulk and surface conductions are important, the MR was found to be negative (*21*). This is usually attributed to the magnetic-field-induced Zeeman energy competing with the spin scattering off the Kondo lattice, effectively reducing the Kondo gap (*21*). Here we investigate surface magnetoresistance at mK temperatures when bulk conductions diminish. Orbital and Zeeman parts of the surface MR can be separately measured in thin plate samples by tiling the magnetic field perpendicular ($\theta = 90°$) and parallel ($\theta = 0°$) to the major surface: MR due to Zeeman effect shows up in both cases while only $\theta = 90°$ configuration will contain the orbital contribution. Fig. 3D summarizes the results from sample S12B. With in-plane field

($\theta = 0°$) at 40 mK the MR $\rho(\theta = 0°)$ due to Zeeman effect is negative and close to quadratic. Little difference was found whether the magnetic field is along or perpendicular to the current direction, confirming the Zeeman nature of in-plane MR. We note such a negative Zeeman contribution to the MR is consistent with the picture of surface Kondo effect discussed above. Tilting the field out of plane, $\rho(\theta = 90°)$ is found to be rather complicated, but we can extract the orbital contribution to the MR using $\Delta\rho = \rho(\theta = 90°) - \rho(\theta = 0°)$. As shown in Fig. 3E, apart from the sharp dip near zero field due to WAL, $\Delta\rho$ is positive and linear for $B > 1T$. The linearity doesn't seem to change when temperature is raised to 350 mK. Linear MR is known to occur in metals with an open Fermi surface (*22*), which is not the case for $SmB_6$ according to recent ARPES results (*8-10*). Disorder or geometry effects could also give rise to linear MR due to symmetric mixing of Hall voltages (*23*). However, in our sample the magnetoresistance $\Delta R_{xx} = \Delta\rho\ L/w$ is 3 times larger than the Hall resistance $R_{xy}$ and makes this scenario unlikely. Interestingly, linear and temperature-independent MR is predicted (*24*) to occur in zero-gap materials with Dirac dispersion in the quantum limit, and is indeed found in Bi-based topological insulators (a review see (*25*)) and in graphene (*26*).

In conclusion, we have verified the existence of a conductive surface state of $SmB_6$ down to 20 mK without signs of localization. The observed WAL effect on the $SmB_6$ surface state is consistent with a TSS with spin-momentum-locking. The temperature dependence of extracted dephasing length indicates that below 100 mK the dephasing of surface electrons is dominated by electron-electron scattering. The saturation of the surface resistivity follows a logarithmic temperature dependence that is best explained by Kondo effect due to the interaction between the surface state and a Kondo lattice with unavoidable defects. In addition, a proper description of the SS of $SmB_6$ needs to be consistent with the observed magnetoresistance that contains a negative Zeeman term and a positive linear orbital term, which may arise as a result of Dirac dispersion.

**Acknowledgements:** This work was supported by UCI CORCL Grant #MIIG-2011-12-8, Sloan Research Fellowship (J.X.) and NSF grant #DMR-0801253. Stimulating discussions with V. Galitski, T. Geballe, H. Jiang, A. Kapitulnik, S. Dodge, S. Kivelson, B. Laughlin, S. Raghu, W. Ho, R. Wu, J Alicea, R. Meng, G. Refael, B. Maple, I. Fisher, E Abrahams, J. Cha, D. Maslov, J. Moore, S. Chakravarty, L. Glazman, R. Greene and B. Rosenow are greatly appreciated.

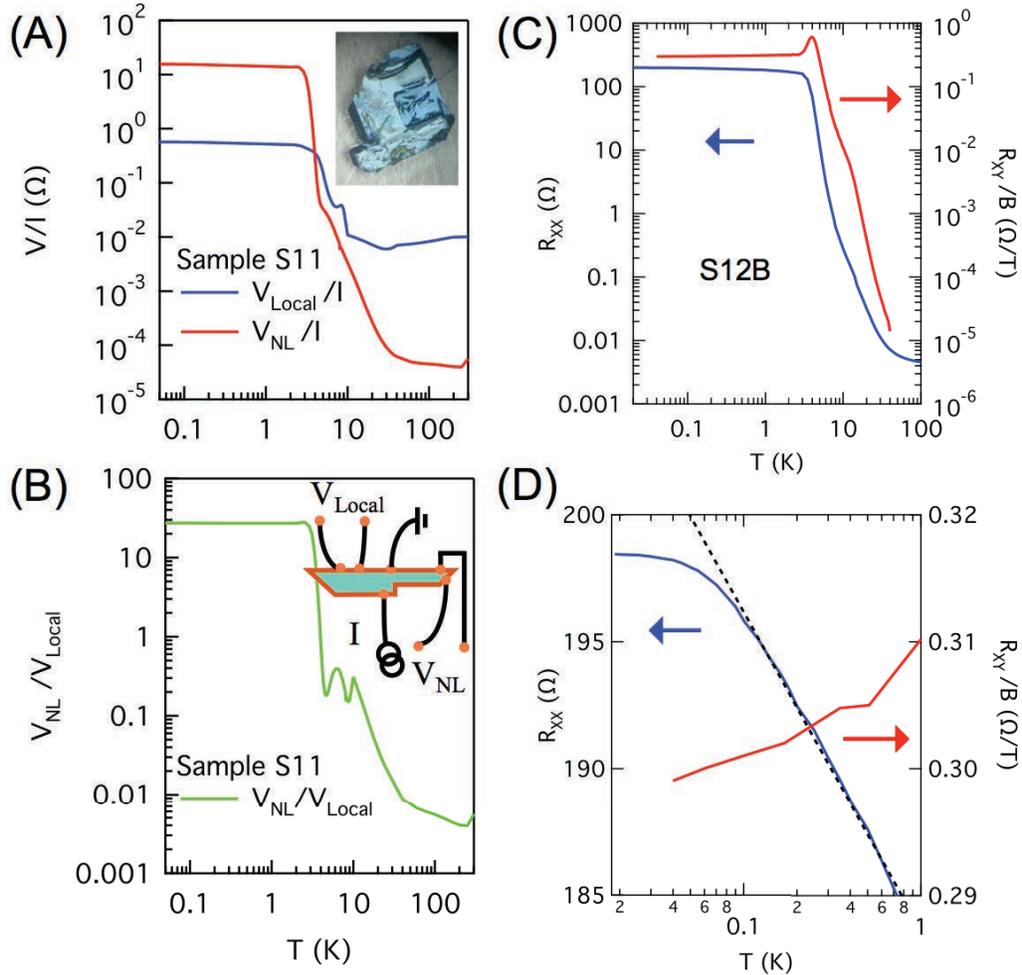

**Fig. 1.** Surface conduction and saturation at mK temperatures. **(A),** Temperature dependence of local and non-local resistance of an irregular thin-plate-shaped *SmB$_6$ crystal* S11 (inset). **(B),** The ratio between non-local and local voltages increases 4 orders of magnitude as the temperature is lowered, and saturates below 3 K. This confirms the domination of surface conduction from 3 K down to 20 mK. Inset is a cross-sectional view of the sample with wiring schematics. **(C),** Typical temperature dependence of $R_{xx}$ and $R_{xy}$/B. **d,** Low temperature behaviors of $R_{xx}$ and $R_{xy}$/B. $R_{xx}$ displays logarithmic temperature dependence between 100 mK and 1K and saturation below 100 mK, reminiscent of Kondo effect.

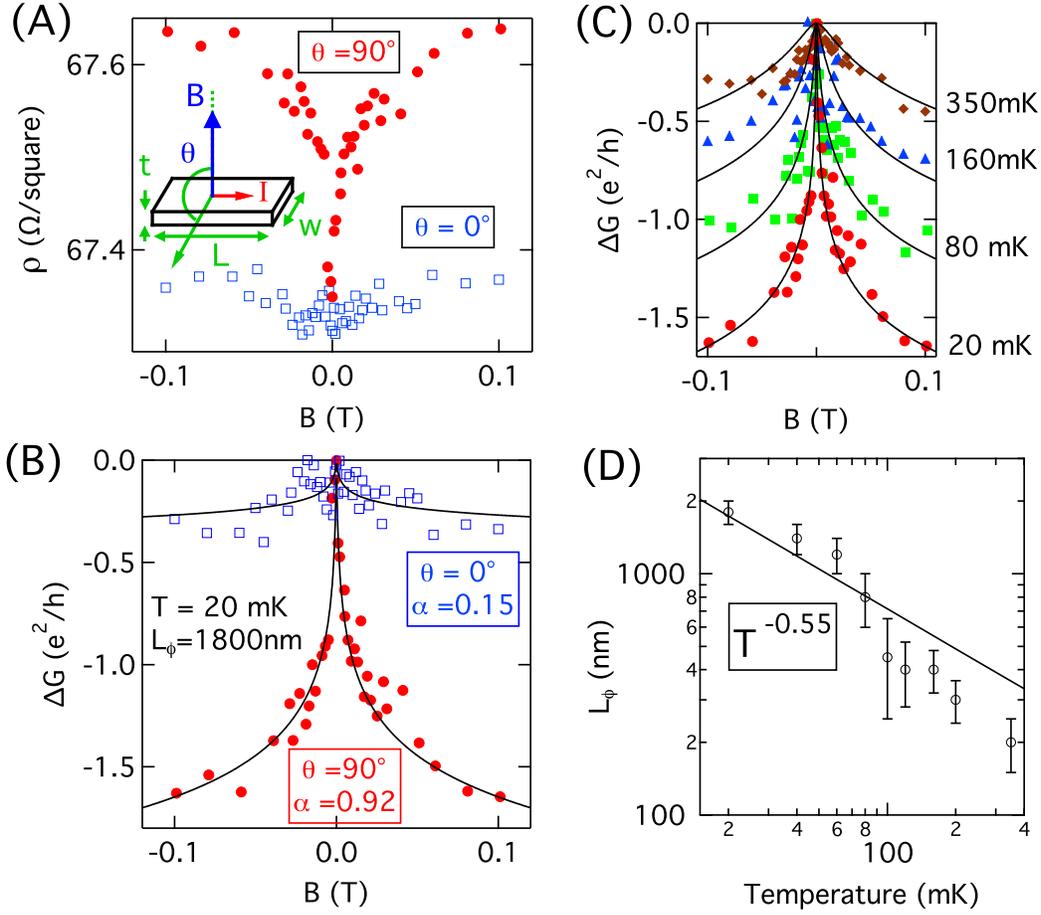

**Fig. 2. Weak antilocalization.** **(A),** Magnetoresistance ρ at 20 mK of a thin plate sample S12B, with perpendicular (red dot) or in-plane (blue square) magnetic fields. Inset, orientation of magnetic field. **(B),** ΔG = Δ(1/ρ) fitted to Hikami-Larkin-Nagaoka (HLN) equation, yielding α = 0.92. **(C),** Temperature evolution of weak Antilocalization effect. Lines are fits using HLN equation with a fixed α = 0.92 and variable dephasing length $L_\phi$. **d,** $L_\phi$ at different temperatures, showing -0.55 power law behavior below 100 mK.

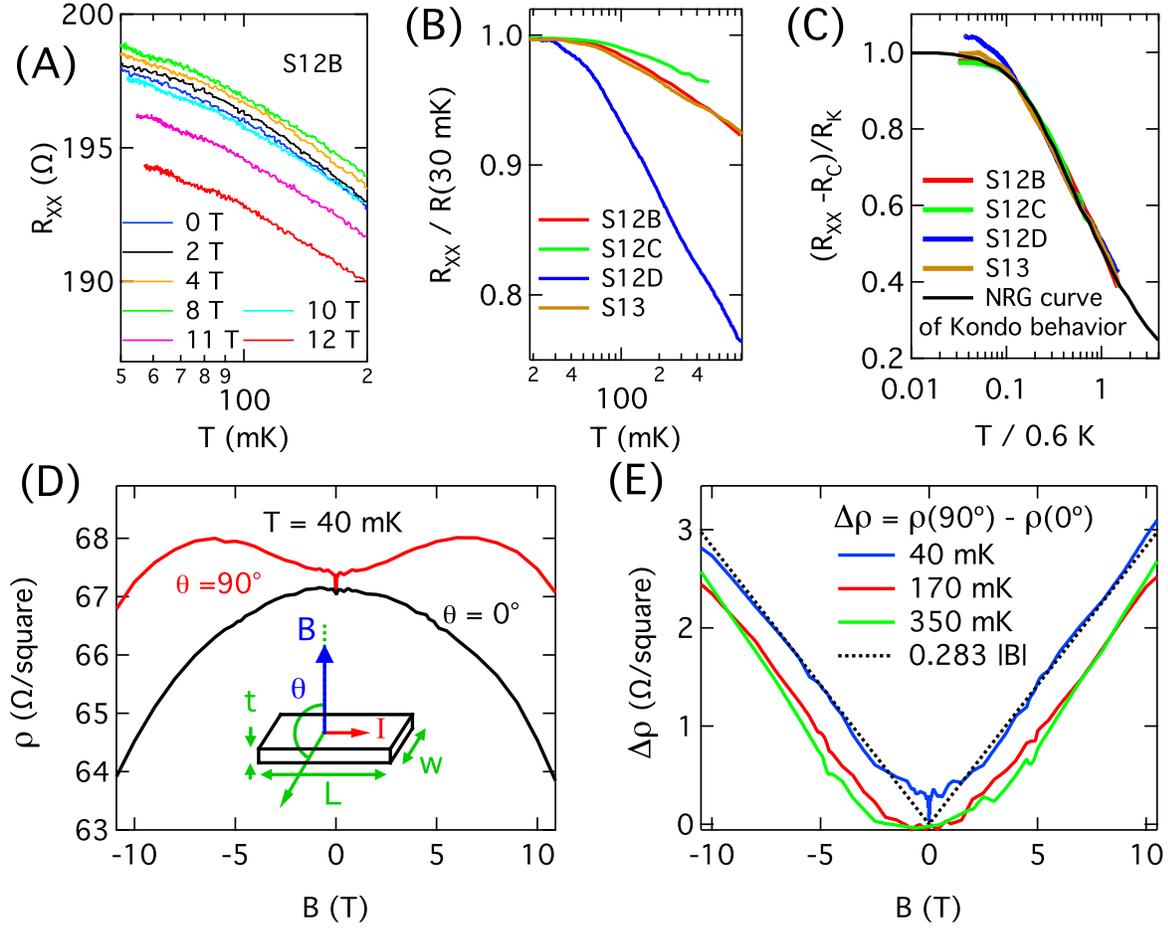

**Fig. 3.** Surface Kondo effect and linear magnetoresistance. **(A),** $R_{xx}$ vs T in sample S12B at various perpendicular fields. **(B),** Temperature dependence of normalized *resistance* for various samples. **(C),** Normalized resistance for various samples in comparison with universal numerical renormalization group (NRG) calculations. **(D),** Magnetoresistance (MR) of sample S12B with perpendicular and in-plane magnetic fields. **(E),** Above 1 Tesla, orbital MR $\Delta\rho = \rho(90°) - \rho(0°)$ shows linear field dependence.

**Supplementary Materials:**

Weak-antilocalization effect has been observed in several samples as illustrated in Fig S1. Fitting them to HLN formula $\Delta G = -\alpha\,[e^2/2\pi^2\hbar][\ln(B_0/B) - \Psi(1/2 + B_0/B)]$, we found that although the dephasing length $L_\phi$ varies among samples, $\alpha$ is almost universal between different samples with the values 0.80, 0.92, 0.95 and 0.97. In particular, after WAL in sample S12B was measured in a first cooldown as described in the main text, we re-prepare the surface and measured it in a second cooldown. In this second cooldown, as shown in Fig S1(A) (B), the dephasing length $L_\phi$ is reduced from 1400 nm to 700 nm at 38 mK, possibly due to a increase amount of surface defects. The parameter $\alpha$ however, only changes slightly from 0.92 to 0.97.

Fig S2 illustrates surface magnetoresistance in sample S14 at 100 mK. Compared to that of S12B (Fig 3 in the main text), both $\rho(\theta = 90°)$ and $\rho(\theta = 0°)$ have qualitatively similar magnetic field dependence. However the exact functional forms differ. For example $\rho(\theta = 0°)$ v.s. B is still negative but is no longer quadratic. Never the less, the orbital part of magnetoresistance $\Delta\rho = \rho(\theta = 90°) - \rho(\theta = 0°)$ is linear with magnetic field below 5 T. The linear coefficient is found to be 0.30 $\Omega/T$, which is quite similar to the 0.283 $\Omega/T$ value found in sample S12B (Fig 3 in the main text). Above 5 T, $\Delta\rho$ starts to saturate, a behavior that awaits further investigation.

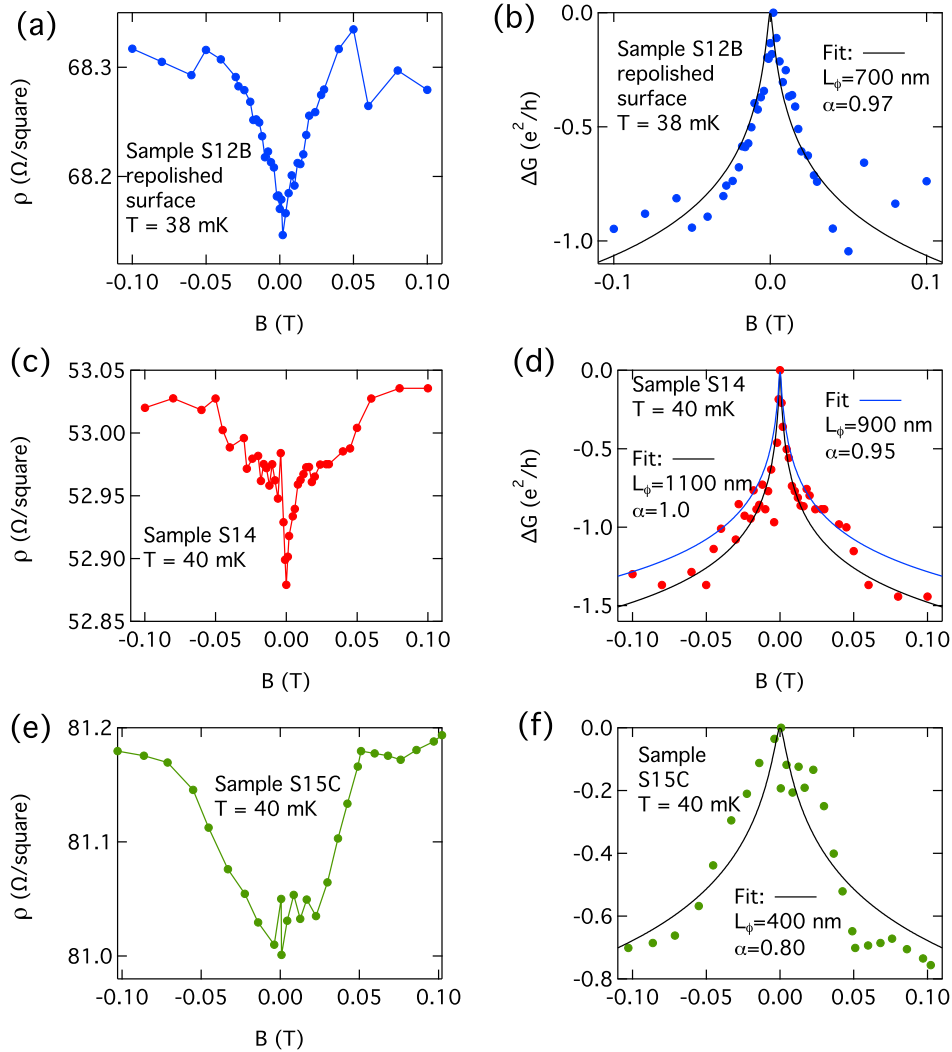

**Fig. S1.** Weak antilocalization in various samples. **(A), (B)** Sample S12B (main text) after repolishing. Repolishing reduces the dephasing length $L_\phi$ to 700 nm at 38 mK. $\alpha$ is however almost unchanged. **(C), (D)** Sample S14, $L_\phi = 900\ nm$, $\alpha = 0.95$ at 40 mK. **(E), (F)** Sample S15C, $L_\phi = 400\ nm$, $\alpha = 0.80$ at 40 mK.

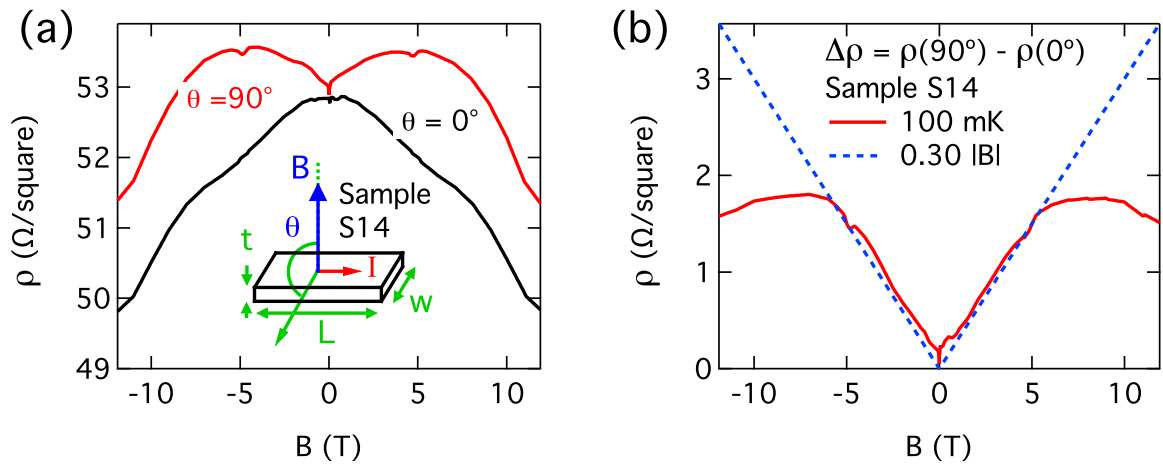

**Fig. S2.** Magnetoresistance for sample S14. **(A),** Magnetoresistance with field in plane (blue) and out-of-plane (red) at 100 mK. The zero field "dip" in $\rho(\theta = 90°)$ is due to WAL. **(B),** The orbital part of surface magnetoresistance $\Delta\rho = \rho(\theta = 90°) - \rho(\theta = 0°)$ is linear with magnetic field when $B < 5\,T$. Blue dashed line is a linear fit $\Delta\rho = 0.30\,|B|$.